\documentclass[prl,twocolumn,showpacs,nofootinbib]{revtex4-2}
\pdfoutput=1

\usepackage{graphicx,amsmath,amssymb,amsthm}
\usepackage[hidelinks]{hyperref}
\usepackage{xcolor}
\usepackage{ulem}
\usepackage{thmtools, thm-restate}
\usepackage{epsfig}
\usepackage{epstopdf}
\usepackage{latexsym}
\usepackage{graphicx}
\usepackage{booktabs}
\usepackage{bbm}
\usepackage{color}
\usepackage{physics}
\usepackage{tensor}
\usepackage{verbatim}
\usepackage[caption=false]{subfig}
\usepackage{tikz}
\usepackage{bigints}
\usepackage{ifthen}
\usetikzlibrary{matrix}
\usetikzlibrary{decorations.markings,calc,shapes,decorations.pathmorphing,arrows.meta}
\usetikzlibrary{patterns}
\usetikzlibrary{positioning}
\usepackage{textpos}

\usepackage{hyperref}

\usepackage{xcolor}

\hypersetup{
colorlinks,
linkcolor={blue!80!black},
citecolor={blue!80!black},
urlcolor={blue!80!black},
linktoc=page
}

\begin{document}
\def\nn{\nonumber}
\def\kc#1{\left(#1\right)}
\def\kd#1{\left[#1\right]}
\def\ke#1{\left\{#1\right\}}
\newcommand\beq{\begin{equation}}
\newcommand\eeq{\end{equation}}
\renewcommand{\Re}{\mathop{\mathrm{Re}}}
\renewcommand{\Im}{\mathop{\mathrm{Im}}}
\renewcommand{\b}[1]{\mathbf{#1}}
\renewcommand{\c}[1]{\mathcal{#1}}
\renewcommand{\u}{\uparrow}
\renewcommand{\d}{\downarrow}
\newcommand{\be}{\begin{equation}}
\newcommand{\ee}{\end{equation}}
\newcommand{\bsigma}{\boldsymbol{\sigma}}
\newcommand{\blambda}{\boldsymbol{\lambda}}
\newcommand{\sgn}{\mathop{\mathrm{sgn}}}
\newcommand{\diag}{\mathop{\mathrm{diag}}}
\newcommand{\Pf}{\mathop{\mathrm{Pf}}}
\newcommand{\half}{{\textstyle\frac{1}{2}}}
\newcommand{\sh}{{\textstyle{\frac{1}{2}}}}
\newcommand{\ish}{{\textstyle{\frac{i}{2}}}}
\newcommand{\thf}{{\textstyle{\frac{3}{2}}}}
\newcommand{\SUN}{SU(\mathcal{N})}
\newcommand{\N}{\mathcal{N}}

\title{Jackiw-Teitelboim Gravity from the Karch-Randall Braneworld}
\preprint{today}
\begin{abstract}

In this letter, we show that Jackiw-Teitelboim (JT) gravity can be naturally realized in the Karch-Randall braneworld. Notably the role of the dilaton in JT gravity is played by the radion in a suitably orbifolded version of the setup. In the classical entanglement entropy calculation, there is an apparent degeneracy of Ryu-Takayanagi surfaces. We demonstrate how quantum fluctuations of the radion/dilaton resolves this would-be classical puzzle regarding {entanglement wedge reconstruction}. 
\end{abstract}
\author{Hao Geng$^{a}$}
\author{Andreas Karch$^{b}$}
\author{Carlos Perez-Pardavila$^{b}$}
\author{Suvrat Raju$^{c}$}
\author{Lisa Randall$^a$}
\author{Marcos Riojas$^{b}$}
\author{Sanjit Shashi$^{b}$}
\affiliation{$^a$Harvard University, 17 Oxford St., Cambridge, MA, 02139, USA.}
\affiliation{$^b$Theory Group, Department of Physics, University of Texas, Austin, TX 78712, USA.}
\affiliation{$^c$International Centre for Theoretical Sciences, Tata Institute of Fundamental Research, Shivakote, Bengaluru 560089, India.}

\maketitle
\section{Introduction}
Recent utilization of the doubly holographic Karch-Randall braneworld \cite{Karch:2000ct,Karch:2000gx} has yielded substantial insights into quantum gravity. These include analytical models that have been used to produce Page curves describing black hole radiation in higher-dimensional (i.e in $d > 2$ spacetime dimensions) setups \cite{Almheiri:2019psy,Geng:2020qvw}\footnote{This is motivated by the earlier work \cite{Penington:2019npb,Almheiri:2019psf,Almheiri:2019hni,Almheiri:2019yqk}.}, demonstrating a connection between the existence of entanglement islands and massive gravity theories \cite{Geng:2020qvw,Geng:2020fxl,Geng:2021hlu}, providing simple holographic models to study quantum field theory in black hole backgrounds \cite{Geng:2021mic,Geng:2022yyy}, and motivating a novel holographic setup---wedge holography \cite{Akal:2020wfl,Miao:2020oey,Geng:2020fxl,Geng:2021hlu}.

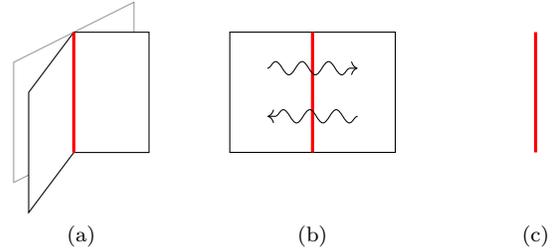
\begin{figure}
\centering
\begin{subfloat}[\label{pic:wedgeHoloBulk}]
{
\begin{tikzpicture}[scale=0.8]
\draw[-,black!40] (-0.5,-0.25) to (1.5,0.75) to (1.5,2.75) to (-0.5,1.75) to (-0.5,-0.25);
\draw[-,fill=white] (0.5,2.25) to (-0.25,1.25) to (-0.25,-0.75) to (0.5,0.25);
\draw[-,fill=white] (0.5,2.25) to (1.75,2.25) to (1.75,0.25) to (0.5,0.25);
\draw[-,red,very thick] (0.5,0.25) to (0.5,2.25);
\node at (2,0) {};
\node at (-0.75,0) {};
\end{tikzpicture}
}
\end{subfloat}\quad\ \ 
\begin{subfloat}[\label{pic:wedgeHoloInter}]
{
\begin{tikzpicture}[scale=0.8,decoration=snake]
\draw[-] (-2.75/2,0.25) to (0,0.25) to (0,2.25) to (-2.75/2,2.25) -- cycle;
\draw[-] (2.75/2,0.25) to (0,0.25) to (0,2.25) to (2.75/2,2.25) -- cycle;
\draw[-,red,very thick] (0,0.25) to (0,2.25);
\node at (0,-0.625) {};

\draw[->,decorate] (-0.75,1.65) to (0.75,1.65);
\draw[->,decorate] (0.75,0.85) to (-0.75,0.85);
\end{tikzpicture}
}
\end{subfloat}\quad\ \ 
\begin{subfloat}[\label{pic:wedgeHoloBound}]
{
\begin{tikzpicture}[scale=0.8]
\draw[-,draw=none] (-2.75/2,0.25) to (0,0.25) to (0,2.25) to (-2.75/2,2.25) -- cycle;
\draw[-,draw=none] (2.75/2,0.25) to (0,0.25) to (0,2.25) to (2.75/2,2.25) -- cycle;
\draw[-,red,very thick] (0,0.25) to (0,2.25);
\node at (0,-0.625) {};
\end{tikzpicture}
}
\end{subfloat}
\caption{\small A schematic representation of the doubly holographic nature of the wedge. (a) is the bulk description, which is just gravity on the $(d+1)$-dimensional wedge with a codimension-two corner in red. (b) is the intermediate description in which the two $d$-dimensional theories share a transparent interface (red). (c) is the boundary description on the interface.}
\label{pic:wedgeHolo}
\end{figure}

Wedge holography is a simple canonical deformation of the original Karch-Randall braneworld which manifests a codimension-two holographic duality \cite{Akal:2020wfl,Miao:2020oey,Geng:2020fxl,Geng:2021hlu}. This setup starts with two AdS$_{d}$ Karch-Randall branes in an ambient AdS$_{d+1}$ with the two branes intersecting each other at the conformal boundary of the ambient spacetime. These branes are treated as ``end-of-the-world" (EOW) branes, meaning that the part of the ambient spacetime behind them is excised. Thus, the leftover bulk spacetime is a wedge (see Fig.~\ref{pic:wedgeHoloBulk}). In analogy with the original Karch-Randall braneworld, the wedge system is \textit{doubly holographic} in that there are three equivalent descriptions related to each other by the AdS/CFT correspondence \cite{Maldacena:1997re,Witten:1998qj,Gubser:1998bc} (see Fig.~\ref{pic:wedgeHolo}):
\begin{itemize}
  \item \textbf{The bulk description:} Einstein-Hilbert gravity in the wedge.
  \item \textbf{The intermediate description:} Two CFT$_d$'s coupled with gravity living on separate AdS$_d$'s and glued to each other along their asymptotic boundaries by imposing transparent boundary conditions.
  \item\textbf{The boundary description:} A CFT$_{d-1}$ living on the common asymptotic boundary of the two branes (which is the corner of the wedge in Fig.~\ref{pic:wedgeHoloBulk}).
\end{itemize}

In this letter, we study wedge holography when the bulk is AdS$_3$ with two AdS$_2$ branes. As opposed to the original wedge holography setup in higher dimensions where the brane fluctuations can be neglected, we will consider fluctuating branes with orbifold projection so that there is only one fluctuating scalar degree of freedom. This scalar is the \textit{radion} in the language of the Randall-Sundrum I setup \cite{Randall:1999ee}. We will derive an action for the radion which is coupled with the 2d localized braneworld gravity. This can be thought of as the low-energy effective theory of the intermediate description. We will see that this theory is precisely the Jackiw-Teitelboim (JT) gravity \cite{Jackiw:1984je,Teitelboim:1983ux}, in which the dilaton field is the radion. 

The importance of our study is that it ensures the consistency of wedge holography in AdS$_3$. This is because when we ignore the brane fluctuations and study the entanglement entropy in AdS$_3$ wedge holography using the Ryu-Takayanagi (RT) formula, we find an infinite number of degenerate RT surfaces. This renders the entanglement wedge, which is an essential element in holography \cite{Dong:2016eik}, as undefined. Taking into account the brane fluctuations, this degeneracy is lifted, as we show explicitly using our effective JT description.

We note that JT gravity \cite{Almheiri:2014cka,Jensen:2016pah,Maldacena:2016upp,Engelsoy:2016xyb} has also played an important role in the recent development of low-dimensional ($d = 2$) quantum gravity \cite{Penington:2019npb,Almheiri:2019psf,Saad:2019lba}. This includes providing an analytically controllable model to compute the Page curve of evaporating black holes \cite{Penington:2019npb,Almheiri:2019psf} and a precise duality between Euclidean JT gravity to a random matrix ensemble \cite{Saad:2019lba}. Our realization of JT gravity in the Karch-Randall braneworld further argues for its significance in our study of quantum gravity \cite{Geng:2022xxx}.

\section{Karch-Randall Braneworld and Wedge Holography}\label{sec:review}
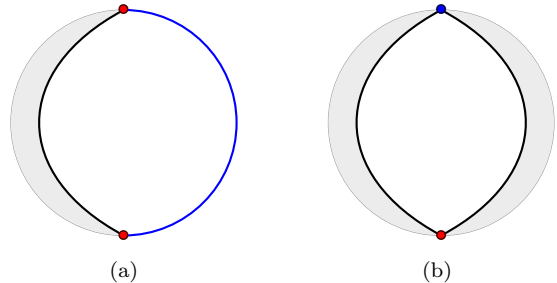
\begin{figure}
\centering
\subfloat[]{
\begin{tikzpicture}[scale=1]

\draw[-,thick,blue] (-3.5,1.5) arc (90:-90:1.5);
\draw[-,black!40] (-3.5,1.5) arc (90:270:1.5);

\draw[-,draw=none,fill=gray!15] (-3.5,1.5) .. controls (-5,0.7) and (-5,-0.7) .. (-3.5,-1.5) arc (-90:-270:1.5);
\draw[-,thick,black] (-3.5,1.5) .. controls (-5,0.7) and (-5,-0.7) .. (-3.5,-1.5);

\node at (-3.5,1.5) {\textcolor{red!100!}{$\bullet$}};
\node at (-3.5,1.5) {\textcolor{black}{$\circ$}};
\node at (-3.5,-1.5) {\textcolor{red!100!}{$\bullet$}};
\node at (-3.5,-1.5) {\textcolor{black}{$\circ$}};
\label{pic:empty1brane}
\end{tikzpicture}
}\qquad\quad
\subfloat[]{
\begin{tikzpicture}[scale=1]
\draw[-,black!40] (0,1.5) arc (90:-90:1.5);
\draw[-,black!40] (0,1.5) arc (90:270:1.5);

\draw[-,draw=none,fill=gray!15] (0,1.5) .. controls (-1.5,0.7) and (-1.5,-0.7) .. (0,-1.5) arc (-90:-270:1.5);
\draw[-,thick,black] (0,1.5) .. controls (-1.5,0.7) and (-1.5,-0.7) .. (0,-1.5);

\draw[-,draw=none,fill=gray!15] (0,1.5) .. controls (1.5,0.7) and (1.5,-0.7) .. (0,-1.5) arc (270:450:1.5);
\draw[-,thick,black] (0,1.5) .. controls (1.5,0.7) and (1.5,-0.7) .. (0,-1.5);

\node at (0,1.5) {\textcolor{blue!100!}{$\bullet$}};
\node at (0,1.5) {\textcolor{black}{$\circ$}};
\node at (0,-1.5) {\textcolor{red!100!}{$\bullet$}};
\node at (0,-1.5) {\textcolor{black}{$\circ$}};
\label{pic:empty2brane}
\end{tikzpicture}}
\caption{\small{(a) A constant time slice of the empty AdS$_{3}$ bulk with one Karch-Randall brane (in black). The brane intersects the asymptotic boundary (in blue) at the two defects (red dots). The holographic dual is a BCFT$_2$ living on the blue interval. (b) The empty AdS$_{3}$ wedge consisting of two KR branes. The two branes intersect at the two defects (red and blue dots). The dual boundary description is hence a conformal quantum mechanics living on the two defects.}}
\label{pic:emptyads3}
\end{figure}

In this section, we briefly review the Karch-Randall braneworld and wedge holography, focusing on the geometric pictures for later holographic study. The original theory of Karch and Randall is described by the following action
\begin{equation}
      \begin{split}
    S=&-\frac{1}{16\pi G_{d+1}}\int d^{d+1}x \sqrt{-g_{\text{bulk}}}(R[g_{\text{bulk}}]-2\Lambda)\\&\qquad\qquad-\frac{1}{8\pi G_{d+1}}\int d^{d}x\sqrt{-h}(K-T)\,,\label{eq:KRaction}
  \end{split}
  \end{equation}
where the first term is the standard Einstein-Hilbert action describing the gravitational in the ambient AdS$_{d+1}$ with a cosmological constant $\Lambda=-\frac{d(d-1)}{2}$,\footnote{We have set the AdS curvature length to one.} the second term describes the embedding of the brane in the ambient AdS$_{d+1}$, $h_{ab}$ is the induced metric on the brane, $K$ is the trace of the extrinsic curvature of the brane and $T$ is the tension of the brane and satisfies $T\leq (d-1)$ as a Karch-Randall brane.\footnote{The critical value $T=d-1$ corresponds to the Randall-Sundrum brane \cite{Randall:1999ee,Randall:1999vf}.} The bulk metric fluctuations satisfy the Neumann boundary condition near the brane
\begin{equation}
  \nabla_{n}\delta g_{\text{bulk $\mu\nu$}}|_{\text{near brane}}=0\,,\label{eq:Neumann}
\end{equation}
where $n$ denotes the normal direction of the brane. With this boundary condition taken into account, setting the variation of the action Equ.~(\ref{eq:KRaction}) to zero we obtain two equations of motion---the usual Einstein's field equation which determines the bulk geometry and the brane-localized equation of motion describing the embedding of the brane in the bulk. The latter is given by \cite{Fujita:2011fp}
\begin{equation}
  K_{ab}=(K-T)h_{ab}\,.\label{eq:braneeom}
\end{equation}
This system of equations can be solved by
\begin{equation}
  ds^{2}_{\text{bulk}}=dr^{2}+\cosh^{2}(r)ds^{2}_{\text{AdS$_d$}}\,,\label{eq:ads}
\end{equation}
where $r\in(-\infty,\infty)$, with the brane given by $r=r_{1}=const.$ and $T=(d-1)\abs{\tanh r_{1}}$ (we only consider positive tension branes).\footnote{If the brane is given by $r=r_{1}<0$ understood as an EOW brane with positive tension the domain of the bulk is $r\in(r_{1},\infty)$ and if the brane is given by $r=r_{2}>0$ understood as an EOW brane with positive tension the bulk domain is $r\in(-\infty,r_{2})$ \cite{Karch:2020iit}.} The bulk metric is just empty AdS$_{d+1}$ written as a foliation into AdS$_{d}$ slices. This is particularly useful in the study of the Karch-Randall braneworld as here the brane geometry $r=const.$ is clearly AdS$_{d}$. In this letter, we focus on the case $d=2$, in which case the geometry can be easily visualized as that on the left of Fig.~\ref{pic:empty1brane}.

In the wedge holography setup \cite{Akal:2020wfl,Miao:2020oey,Geng:2020fxl,Geng:2021hlu}, we add another (positive tension) Karch-Randall brane. This system is described by the following action
\begin{equation}
\begin{split}
S =& -\frac{1}{16\pi G_{3}}\int d^{3}x \sqrt{-g_{\text{bulk}}}(R[g_{\text{bulk}}]-2\Lambda)\\&\qquad\qquad-\frac{1}{8\pi G_{3}}\int_{i=1,2} d^{d}x\sqrt{-h_{i}}(K_{i}-T_{i})\,,\label{eq:wedgeaction}
\end{split}
\end{equation}
where we impose the Neumann boundary conditions Equ.~(\ref{eq:Neumann}) for the bulk metric fluctuation near both branes. The equations of motion still consist of two parts---the bulk Einstein's equation and the two brane-localized equations, with each one the same as Equ.~(\ref{eq:braneeom}) but with their own tensions $T_{1}$ and $T_{2}$. Hence, a solution is given by the same bulk metric Equ.~(\ref{eq:ads}) with $d=2$ where $r\in(r_{1},r_{2})$, $T_{1}=\tanh(-r_{1})$, $T_{2}=\tanh r_{2}$, $r_{1}<0$ and $r_{2}>0$. This geometry is visualized in Fig.~\ref{pic:empty2brane}.

\section{Realizing JT Gravity}\label{sec:JT}
In the original wedge holography papers \cite{Akal:2020wfl,Miao:2020oey,Geng:2020fxl,Geng:2021hlu} the branes are treated as rigid. This assumption is rooted in the original Karch-Randall setup where we just have one brane and we can take the brane to be sitting at an orbifold fixed point to model an EOW brane in analog to O-planes in string theory \cite{Polchinski:1998rq,Johnson:2003gi}. However, reminding ourselves of the Randall-Sundrum I (RS1) setup \cite{Randall:1999ee} with two RS branes, the relative distance between the branes is a dynamical variable called the radion, which has important phenomenological implications \cite{ArkaniHamed:2000ds}. Hence in the case of wedge holography, we should not neglect the possibility of brane fluctuations.

In this section, we consider small brane fluctuations (relative to the bulk AdS$_3$ curvature scale) and derive the effective action for the coupling between these small fluctuations and the brane-localized graviton mode by doing dimensional reduction. While brane fluctuations can be neglected in higher dimensions, we will see that this small contribution plays an important role when the bulk is AdS$_3$. The detailed derivation and a more careful analysis will be presented in \cite{Geng:2022xxx}.

To be precise, we will consider two branes with tensions $T_{1}$ and $T_{2}$ sitting at $r=r_{1}+\delta\phi_{1}(x)$ and $r=r_{2}+\delta\phi_{2}(x)$ slices of the following bulk geometry
\begin{equation}
ds^{2}=dr^{2}+\cosh^{2}(r) g_{ab}(x) dx^{a}dx^{b}\,,\label{eq:ansatz}
\end{equation}
with the relationships $T_{1}=-\tanh r_{1}>0$ and $T_{2}=\tanh r_{2}>0$ satisfied and $\delta\phi_{1},\delta\phi_{2}\ll1$ (the AdS length scale has been set to one). The wedge runs from $r=r_{1}+\delta\phi_{1}(x)$ to $r=r_{2}+\delta\phi_{2}(x)$, and the metric ansatz Equ.~(\ref{eq:ansatz}) describes the lowest 2d graviton mode (the higher modes would depend on the coordinate $r$ as $g_{ab}(r,x)$).

To obtain the low-energy effective action, we have to plug the above ansatz of the metric and brane locations into Equ.~(\ref{eq:wedgeaction}) and keep terms only up to order $\mathcal{O}(\delta\phi^{2})$. This yields the following action\footnote{For details of the derivation, we refer the readers to \cite{Geng:2022xxx}.}
\begin{equation}
  S_{\text{eff}}=S_{0}-\frac{1}{16\pi G_{3}}\int d^{2}x\sqrt{-g}\phi(x)\Big[R[g]+2\Big]+S_{\text{dilaton}}\,,
\end{equation}
where $\phi(x)=\delta\phi_{2}(x)-\delta\phi_{1}(x)$ is the radion mode, $S_{0}$ is a purely topological term
\begin{equation}
  \begin{split}
    S_{0}&=-\frac{r_{2}-r_{1}}{16\pi G_{3}}\int d^{2}x\sqrt{-g}R[g]\,,
  \end{split}
\end{equation}
and $S_{\text{dilaton}}$ is the kinetic term for the brane fluctuations
\begin{equation}
\begin{split}
S_{\text{dilaton}}=&-\frac{1}{8\pi G_{3}}\int d^{2}x\sqrt{-g}\Big[\frac{T_{2}}{2}\nabla_{\mu}\delta\phi_{2}\nabla^{\mu}\delta\phi_{2}+T_{2}(\delta\phi_{2})^2\\
&\qquad\qquad\quad\quad\quad+\frac{T_{1}}{2}\nabla_{\mu}\delta\phi_{1}\nabla^{\mu}\delta\phi_{1}+T_{1}(\delta\phi_{1})^2\Big]\,.\label{eq:dilaton2d}
\end{split}
\end{equation}
In analogy to the Randall-Sundrum scenario, to isolate the dynamics of the radion we have to orbifold our setup. We can project out either $\delta\phi_{1}(x)$ or $\delta\phi_{2}(x)$ by a canonical orbifold prescription: start with an array of an infinite number of paired wedges, mod out the translation symmetry on the array to a single pair of wedges with $\mathbb{Z}_{2}$ symmetry, and then mod out \textit{this} $\mathbb{Z}_2$ symmetry (see \cite{Geng:2022xxx} for details) to get a single wedge with only one fluctuating brane. The possibility of projecting out either $\delta\phi_1(x)$ or $\delta\phi_2(x)$ are gauge-equivalent. To see this gauge equivalence, without loss of generality we can project out $\delta\phi_1(x)$ and redefine $g_{ab}(x)$ such that the final action is independent of $r_1$ and $r_2$. Specifically, by setting
\begin{equation}
\delta\phi_{1}(x)=0\,,\quad g_{ab}(x)\rightarrow e^{-T_{2}\delta\phi_{2}(x)}g_{ab}(x)\,,
\end{equation}
the total action becomes
\begin{equation}
  S_{\text{eff}}=S_{0}-\frac{1}{16\pi G_{3}}\int d^{2}x \sqrt{-g}\phi(x)\Bigg[R[g]+2\Bigg]\,,\label{eq:JTbulk}
\end{equation}
where now the dilaton field is $\phi(x)=\delta\phi_{2}(x)$. This is precisely Jackiw-Teitelboim gravity \cite{Jackiw:1984je,Teitelboim:1983ux} for AdS$_2$.

However, there is no nontrivial profile for the dilaton on the AdS$_2$ background without a cutoff \cite{Maldacena:1998uz}. This cutoff breaks the asymptotic conformal symmetry and allows the theory to have finite energy excitations \cite{Maldacena:2016upp}. Hence we fix a small cutoff $\epsilon$ and fix the boundary metric for the nearly-AdS$_2$ and the dilaton field to be
\begin{equation}
  g_{ab}(x)|_{\text{bdy}}dx^{a}dx^{b}=-\frac{du^2}{\epsilon^2}\,,\ \ \phi(x)|_{\text{bdy}}=\phi_{b}=\frac{\phi_{r}}{\epsilon}\,.
\end{equation}
A subtlety of our derivation is that we have to take $\phi_{b} \sim \epsilon$ (or $\phi_{r}\sim\epsilon^2$) to ensure the brane fluctuation is always small, i.e. $\delta\phi_{i}(x)\ll1$.

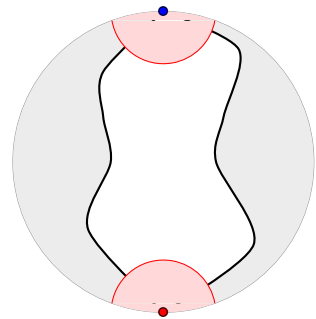
\begin{figure}
\centering
\begin{tikzpicture}
\draw[-,black!40] (0,2) arc (90:-90:2);
\draw[-,black!40] (0,2) arc (90:270:2);
\draw[-,draw=none,fill=gray!15] plot [smooth,tension=0.5] coordinates{(0,2)(-0.8,1.2)(-0.8,0.6)(-0.7,0)(-1,-0.9)(0,-2)} (0,-2) arc (-90:-270:2);
\draw[thick] plot [smooth,tension=0.5] coordinates{(0,2)(-0.8,1.2)(-0.8,0.6)(-0.7,0)(-1,-0.9)(0,-2)};
\draw[-,draw=none,fill=gray!15] plot [smooth,tension=0.5] coordinates{(0,2)(1,1.5)(0.8,0.6)(0.7,0)(1.2,-1.1)(0,-2)} (0,-2) arc (270:450:2);
\draw[thick] plot [smooth,tension=0.5] coordinates{(0,2)(1,1.5)(0.8,0.6)(0.7,0)(1.2,-1.1)(0,-2)};
\draw[-,draw=none,fill=red!15] (-0.684,-1.88) arc (170:10:0.6946)--(-0.684,-1.88) arc (-110:-70:2);
\draw[-,draw=none,fill=red!15] (-0.684,1.88) arc (-170:-10:0.6946)--(-0.684,1.88) arc (110:70:2);
\draw[-,red] (-0.684,1.88) arc (-170:-10:0.6946);
\draw[-,red] (-0.684,-1.88) arc (170:10:0.6946);
\node at (0,2) {\textcolor{blue!100!}{$\bullet$}};
\node at (0,2) {\textcolor{black}{$\circ$}};
\node at (0,-2) {\textcolor{red!100!}{$\bullet$}};
\node at (0,-2) {\textcolor{black}{$\circ$}};
\end{tikzpicture}
\caption{\small The shapes of the two branes are fluctuating (though more rigorously we have also to impose the orbifold projection). The AdS$_2$ cutoff is induced by the AdS$_3$ cutoff (the red curve). The shaded regions are excised. Now the regulated asymptotic boundary of the wedge are two intervals.}
\label{pic:3dcutoff}
\end{figure}

Translating to the AdS$_{3}$ language Equ.~(\ref{eq:ansatz}), we have to take a cutoff of the AdS$_{3}$ wedge and fix the metric there to be (see Fig.~\ref{pic:3dcutoff} for an indication)
\begin{equation}
  ds^{2}_{\text{bdy}}=dr^{2}-\cosh^{2}(r)\frac{du^2}{\epsilon^2}\,.
\end{equation}

\noindent Now that we have identified this cutoff and boundary conditions for the metric in the AdS$_3$ bulk, we have to add the corresponding boundary term to the total action Equ.~(\ref{eq:wedgeaction}). The boundary term is the usual Gibbons-Hawking term
\begin{equation}
  S_{\text{bdy}}=S_{\text{GH}}=-\frac{1}{8\pi G_{3}}\int d^{2}x_{\text{bdy}}\sqrt{-h^{\text{bdy}}}K^{(3)}\,,
\end{equation}
which can be shown to be
\begin{equation}
\begin{split}
S_{\text{bdy}} &= -\frac{2(r_{2}-r_{1})}{16\pi G_{3}}\int_{\partial\mathcal{M}} dx\sqrt{-h}K-\frac{2\phi_{b}}{16\pi G_{3}}\int_{\partial\mathcal{M}}dx\sqrt{-h}K\,,
\end{split}
\end{equation}
where $K$ is the trace of the extrinsic curvature of the cutoff boundary of the AdS$_2$. With this term added to Equ.~(\ref{eq:JTbulk}), we get the full-fledged JT gravity action as that in \cite{Harlow:2018tqv}. For more details we refer the readers to \cite{Geng:2022xxx}.

We emphasize that the scaling $\phi_{b}\sim\epsilon$ is special in our setup. For example, it does not appear in dimensional reduction for the near-horizon region of four-dimensional near-extremal black holes \cite{Nayak:2018qej,Yang:2018gdb}. This indicates that our theory has a different UV completion than previously studied. In our case, the UV completion is a conformal quantum mechanics system \cite{Geng:2022xxx}.

\section{A Puzzle and Its Resolution}\label{sec:paradox}

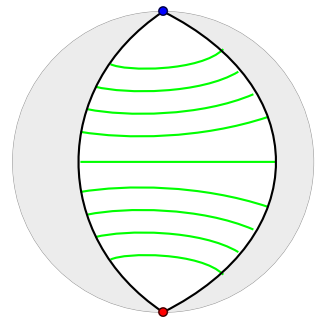
\begin{figure}
\centering
\begin{tikzpicture}
\draw[-,black!40] (0,2) arc (90:-90:2);
\draw[-,black!40] (0,2) arc (90:270:2);

\draw[-,draw=none,fill=gray!15] (0,2) .. controls (-1.5,1) and (-1.5,-1) .. (0,-2) arc (-90:-270:2);

\draw[-,draw=none,fill=gray!15] (0,2) .. controls (2,1) and (2,-1) .. (0,-2) arc (270:450:2);

\draw[-,thick,green] (0.8,1.5)..controls (0.5,1.2) and (-0.5,1.2)..(-0.7,1.3);
\draw[-,thick,green] (1,1.2)..controls (0.5,0.9) and (-0.5,0.9)..(-0.9,1);
\draw[-,thick,green] (1.2,0.9)..controls (0.5,0.6) and (-0.5,0.6)..(-1,0.7);
\draw[-,thick,green] (1.4,0.6)..controls (0.5,0.3) and (-0.5,0.3)..(-1.1,0.4);
\draw[-,thick,green] (1.5,0) to (-1.1,0);
\draw[-,thick,green] (0.8,-1.5)..controls (0.5,-1.2) and (-0.5,-1.2)..(-0.7,-1.3);
\draw[-,thick,green] (1,-1.2)..controls (0.5,-0.9) and (-0.5,-0.9)..(-0.9,-1);
\draw[-,thick,green] (1.2,-0.9)..controls (0.5,-0.6) and (-0.5,-0.6)..(-1,-0.7);
\draw[-,thick,green] (1.4,-0.6)..controls (0.5,-0.3) and (-0.5,-0.3)..(-1.1,-0.4);

\draw[-,thick,black] (0,2) .. controls (-1.5,1) and (-1.5,-1) .. (0,-2);
\draw[-,thick,black] (0,2) .. controls (2,1) and (2,-1) .. (0,-2);

\node at (0,2) {\textcolor{blue!100!}{$\bullet$}};
\node at (0,2) {\textcolor{black}{$\circ$}};
\node at (0,-2) {\textcolor{red!100!}{$\bullet$}};
\node at (0,-2) {\textcolor{black}{$\circ$}};

\end{tikzpicture}
\caption{\small The empty AdS$_{3}$ wedge consisting of two KR branes. The bulk geometry is the AdS$_3$ global patch. The green curves are representatives of the degenerate RT surfaces calculating the entanglement entropy between the two defects (the blue dot and the red dot).}
\label{pic:emptyads3rig}
\end{figure}

A canonical quantity to study in AdS$_{3}$ wedge holography is the entanglement entropy between the two asymptotic defects which support the dual quantum mechanics (see Fig.~\ref{pic:empty2brane}). This can be calculated using the Ryu-Takayanagi (RT) formula \cite{Ryu:2006bv,Ryu:2006ef} by studying the RT surfaces that connect the two branes \cite{Akal:2020wfl,Geng:2020fxl}. Choosing global AdS$_2$ for those AdS$_2$ slices, the bulk metric Equ.~(\ref{eq:ads}) can be written as
\begin{equation}
ds^2=dr^2+\cosh^{2}(r)\big[-\cosh^2(\eta)d\tau^2+d\eta^2\big]\,,\label{eq:metric3}
\end{equation}
if we ignore the brane fluctuations $\delta\phi_{i}(x)$. The RT surface can be parametrized as $\eta=\eta(r)$, $\tau=const.$ which leads to the area functional
\begin{equation}
A=\int _{r_{1}}^{r_{2}}dr\sqrt{1+\cosh^{2}(r)\eta'(r)^2}\,.
\end{equation}
It is easy to show that, with the boundary term carefully taken into account, the variation of the area functional vanishes if and only if $\eta'(r)=0$. Therefore, any constant $\eta$ surface is an RT surface and they all have equal area $A=r_{2}-r_{1}$ (see Fig.~\ref{pic:emptyads3rig} for an illustration). According to this result, there would be no definite RT surface for the entanglement entropy between the two defects and as a corollary there would be no definite entanglement wedge for either of the defects. This would be problematic because it would preclude a holographic interpretation from the point of view of bulk (or more precisely, entanglement wedge \cite{Jafferis:2015del}) reconstruction.

Fortunately, this puzzling result occurs only when the branes are treated as rigid. We expect brane fluctuations to lift the degeneracy of the RT surfaces. Because the brane fluctuation is in general time-dependent, we should use the HRT formula \cite{Hubeny:2007xt} to compute the entanglement entropy from the wedge. However, this would be complicated as the result is heavily dependent on properly treating the boundary condition near the brane.

This challenge of computing entanglement entropy with brane fluctuations is readily met if we instead do it using our effective braneworld theory Equ.~(\ref{eq:JTbulk}). In that case we are just computing the entanglement entropy between the two cutoff boundaries of the nearly-AdS$_2$ space. By solving the equation of motion of the dilaton and 2d metric, we have that
\begin{align}
ds^{2}_{2d}&=-\cosh^{2}(\eta)d\tau^2+d\eta^2\,,\\
\phi(x)&=\phi_{r}\cosh\eta \cos\tau\,,
\end{align}
where $\phi_{r}$ is a constant. Thus, using the maxmin prescription \cite{Wall:2012uf} of the HRT formula we have
\begin{equation}
S_{\text{EE}}=\max_{\tau} \min_{\eta}\Big(\frac{\phi_{0}+\phi(x)}{4G_{3}}\Big)=\frac{\phi_{0}+\phi_{r}}{4G_{3}}\,,
\end{equation}
where the minimization has a solution $\eta=0$, $\tau=0$. As a result, from the bulk AdS$_3$ perspective Equ.~(\ref{eq:metric3}), the RT surface is just the $\eta=0$ surface which sits precisely in the middle of the bulk (i.e. the middle green surface in Fig.~\ref{pic:emptyads3rig}). Thus, the entanglement wedge of either defect is the half bulk region next to it, which is consistent with the symmetry between the two defects.

\section{Conclusions}\label{sec:conclusion}
In this letter, we studied wedge holography when the bulk is three-dimensional. We found that it is necessary to consider brane fluctuations, identified as the radion mode in the braneworld language \cite{Randall:1999ee,Randall:1999vf}, for the apparent consistency of the holographic setup. We found that the low-energy dynamics of the brane fluctuations with orbifold projection is precisely Jackiw-Teitelboim gravity.

\section*{Acknowledgements}
We are grateful to Andreas Blommaert, Kristan Jensen, Hong Liu and Douglas Stanford for useful discussions. HG is very grateful to his parents and recommenders. The work of HG is supported by the grant (272268) from the Moore Foundation ``Fundamental Physics from Astronomy and Cosmology." The work of AK and MR was supported, in part, by a grant from the Simons Foundation (Grant 651440, AK). The work of CP was supported in part by the National Science Foundation under Grant No.~PHY-1914679 and by the Robert N. Little Fellowship. The work of SR was partially supported by a Swarnajayanti fellowship, DST/SJF/PSA-02/2016-17 from the Department of Science and Technology (India). Research at ICTS-TIFR is supported by the government of India through the Department of Atomic Energy grant RTI4001. The work of LR is supported by NSF grants PHY-1620806 and PHY-1915071, the Chau Foundation HS Chau postdoc award, the Kavli Foundation grant ``Kavli Dream Team,'' and the Moore Foundation Award 8342. SS is supported by NSF Grants No. PHY-1914679 and PHY-2112725. 

\bibliographystyle{apsrev4-1}
\bibliography{main}

\begin{thebibliography}{42}%
\makeatletter
\providecommand \@ifxundefined [1]{%
 \@ifx{#1\undefined}
}%
\providecommand \@ifnum [1]{%
 \ifnum #1\expandafter \@firstoftwo
 \else \expandafter \@secondoftwo
 \fi
}%
\providecommand \@ifx [1]{%
 \ifx #1\expandafter \@firstoftwo
 \else \expandafter \@secondoftwo
 \fi
}%
\providecommand \natexlab [1]{#1}%
\providecommand \enquote  [1]{``#1''}%
\providecommand \bibnamefont  [1]{#1}%
\providecommand \bibfnamefont [1]{#1}%
\providecommand \citenamefont [1]{#1}%
\providecommand \href@noop [0]{\@secondoftwo}%
\providecommand \href [0]{\begingroup \@sanitize@url \@href}%
\providecommand \@href[1]{\@@startlink{#1}\@@href}%
\providecommand \@@href[1]{\endgroup#1\@@endlink}%
\providecommand \@sanitize@url [0]{\catcode `\\12\catcode `\$12\catcode
  `\&12\catcode `\#12\catcode `\^12\catcode `\_12\catcode `\%12\relax}%
\providecommand \@@startlink[1]{}%
\providecommand \@@endlink[0]{}%
\providecommand \url  [0]{\begingroup\@sanitize@url \@url }%
\providecommand \@url [1]{\endgroup\@href {#1}{\urlprefix }}%
\providecommand \urlprefix  [0]{URL }%
\providecommand \Eprint [0]{\href }%
\providecommand \doibase [0]{http://dx.doi.org/}%
\providecommand \selectlanguage [0]{\@gobble}%
\providecommand \bibinfo  [0]{\@secondoftwo}%
\providecommand \bibfield  [0]{\@secondoftwo}%
\providecommand \translation [1]{[#1]}%
\providecommand \BibitemOpen [0]{}%
\providecommand \bibitemStop [0]{}%
\providecommand \bibitemNoStop [0]{.\EOS\space}%
\providecommand \EOS [0]{\spacefactor3000\relax}%
\providecommand \BibitemShut  [1]{\csname bibitem#1\endcsname}%
\let\auto@bib@innerbib\@empty
\bibitem [{\citenamefont {Karch}\ and\ \citenamefont
  {Randall}(2001{\natexlab{a}})}]{Karch:2000ct}%
  \BibitemOpen
  \bibfield  {author} {\bibinfo {author} {\bibfnamefont {A.}~\bibnamefont
  {Karch}}\ and\ \bibinfo {author} {\bibfnamefont {L.}~\bibnamefont
  {Randall}},\ }\href {\doibase 10.1088/1126-6708/2001/05/008} {\bibfield
  {journal} {\bibinfo  {journal} {JHEP}\ }\textbf {\bibinfo {volume} {05}},\
  \bibinfo {pages} {008} (\bibinfo {year} {2001}{\natexlab{a}})},\ \Eprint
  {http://arxiv.org/abs/hep-th/0011156} {arXiv:hep-th/0011156} \BibitemShut
  {NoStop}%
\bibitem [{\citenamefont {Karch}\ and\ \citenamefont
  {Randall}(2001{\natexlab{b}})}]{Karch:2000gx}%
  \BibitemOpen
  \bibfield  {author} {\bibinfo {author} {\bibfnamefont {A.}~\bibnamefont
  {Karch}}\ and\ \bibinfo {author} {\bibfnamefont {L.}~\bibnamefont
  {Randall}},\ }\href {\doibase 10.1088/1126-6708/2001/06/063} {\bibfield
  {journal} {\bibinfo  {journal} {JHEP}\ }\textbf {\bibinfo {volume} {06}},\
  \bibinfo {pages} {063} (\bibinfo {year} {2001}{\natexlab{b}})},\ \Eprint
  {http://arxiv.org/abs/hep-th/0105132} {arXiv:hep-th/0105132} \BibitemShut
  {NoStop}%
\bibitem [{\citenamefont {Almheiri}\ \emph
  {et~al.}(2020{\natexlab{a}})\citenamefont {Almheiri}, \citenamefont
  {Mahajan},\ and\ \citenamefont {Santos}}]{Almheiri:2019psy}%
  \BibitemOpen
  \bibfield  {author} {\bibinfo {author} {\bibfnamefont {A.}~\bibnamefont
  {Almheiri}}, \bibinfo {author} {\bibfnamefont {R.}~\bibnamefont {Mahajan}}, \
  and\ \bibinfo {author} {\bibfnamefont {J.~E.}\ \bibnamefont {Santos}},\
  }\href {\doibase 10.21468/SciPostPhys.9.1.001} {\bibfield  {journal}
  {\bibinfo  {journal} {SciPost Phys.}\ }\textbf {\bibinfo {volume} {9}},\
  \bibinfo {pages} {001} (\bibinfo {year} {2020}{\natexlab{a}})},\ \Eprint
  {http://arxiv.org/abs/1911.09666} {arXiv:1911.09666 [hep-th]} \BibitemShut
  {NoStop}%
\bibitem [{\citenamefont {Geng}\ and\ \citenamefont
  {Karch}(2020)}]{Geng:2020qvw}%
  \BibitemOpen
  \bibfield  {author} {\bibinfo {author} {\bibfnamefont {H.}~\bibnamefont
  {Geng}}\ and\ \bibinfo {author} {\bibfnamefont {A.}~\bibnamefont {Karch}},\
  }\href {\doibase 10.1007/JHEP09(2020)121} {\bibfield  {journal} {\bibinfo
  {journal} {JHEP}\ }\textbf {\bibinfo {volume} {09}},\ \bibinfo {pages} {121}
  (\bibinfo {year} {2020})},\ \Eprint {http://arxiv.org/abs/2006.02438}
  {arXiv:2006.02438 [hep-th]} \BibitemShut {NoStop}%
\bibitem [{\citenamefont {Penington}(2020)}]{Penington:2019npb}%
  \BibitemOpen
  \bibfield  {author} {\bibinfo {author} {\bibfnamefont {G.}~\bibnamefont
  {Penington}},\ }\href {\doibase 10.1007/JHEP09(2020)002} {\bibfield
  {journal} {\bibinfo  {journal} {JHEP}\ }\textbf {\bibinfo {volume} {09}},\
  \bibinfo {pages} {002} (\bibinfo {year} {2020})},\ \Eprint
  {http://arxiv.org/abs/1905.08255} {arXiv:1905.08255 [hep-th]} \BibitemShut
  {NoStop}%
\bibitem [{\citenamefont {Almheiri}\ \emph
  {et~al.}(2019{\natexlab{a}})\citenamefont {Almheiri}, \citenamefont
  {Engelhardt}, \citenamefont {Marolf},\ and\ \citenamefont
  {Maxfield}}]{Almheiri:2019psf}%
  \BibitemOpen
  \bibfield  {author} {\bibinfo {author} {\bibfnamefont {A.}~\bibnamefont
  {Almheiri}}, \bibinfo {author} {\bibfnamefont {N.}~\bibnamefont
  {Engelhardt}}, \bibinfo {author} {\bibfnamefont {D.}~\bibnamefont {Marolf}},
  \ and\ \bibinfo {author} {\bibfnamefont {H.}~\bibnamefont {Maxfield}},\
  }\href {\doibase 10.1007/JHEP12(2019)063} {\bibfield  {journal} {\bibinfo
  {journal} {JHEP}\ }\textbf {\bibinfo {volume} {12}},\ \bibinfo {pages} {063}
  (\bibinfo {year} {2019}{\natexlab{a}})},\ \Eprint
  {http://arxiv.org/abs/1905.08762} {arXiv:1905.08762 [hep-th]} \BibitemShut
  {NoStop}%
\bibitem [{\citenamefont {Almheiri}\ \emph
  {et~al.}(2020{\natexlab{b}})\citenamefont {Almheiri}, \citenamefont
  {Mahajan}, \citenamefont {Maldacena},\ and\ \citenamefont
  {Zhao}}]{Almheiri:2019hni}%
  \BibitemOpen
  \bibfield  {author} {\bibinfo {author} {\bibfnamefont {A.}~\bibnamefont
  {Almheiri}}, \bibinfo {author} {\bibfnamefont {R.}~\bibnamefont {Mahajan}},
  \bibinfo {author} {\bibfnamefont {J.}~\bibnamefont {Maldacena}}, \ and\
  \bibinfo {author} {\bibfnamefont {Y.}~\bibnamefont {Zhao}},\ }\href {\doibase
  10.1007/JHEP03(2020)149} {\bibfield  {journal} {\bibinfo  {journal} {JHEP}\
  }\textbf {\bibinfo {volume} {03}},\ \bibinfo {pages} {149} (\bibinfo {year}
  {2020}{\natexlab{b}})},\ \Eprint {http://arxiv.org/abs/1908.10996}
  {arXiv:1908.10996 [hep-th]} \BibitemShut {NoStop}%
\bibitem [{\citenamefont {Almheiri}\ \emph
  {et~al.}(2019{\natexlab{b}})\citenamefont {Almheiri}, \citenamefont
  {Mahajan},\ and\ \citenamefont {Maldacena}}]{Almheiri:2019yqk}%
  \BibitemOpen
  \bibfield  {author} {\bibinfo {author} {\bibfnamefont {A.}~\bibnamefont
  {Almheiri}}, \bibinfo {author} {\bibfnamefont {R.}~\bibnamefont {Mahajan}}, \
  and\ \bibinfo {author} {\bibfnamefont {J.}~\bibnamefont {Maldacena}},\
  }\href@noop {} {\  (\bibinfo {year} {2019}{\natexlab{b}})},\ \Eprint
  {http://arxiv.org/abs/1910.11077} {arXiv:1910.11077 [hep-th]} \BibitemShut
  {NoStop}%
\bibitem [{\citenamefont {Geng}\ \emph {et~al.}(2020)\citenamefont {Geng},
  \citenamefont {Karch}, \citenamefont {Perez-Pardavila}, \citenamefont {Raju},
  \citenamefont {Randall}, \citenamefont {Riojas},\ and\ \citenamefont
  {Shashi}}]{Geng:2020fxl}%
  \BibitemOpen
  \bibfield  {author} {\bibinfo {author} {\bibfnamefont {H.}~\bibnamefont
  {Geng}}, \bibinfo {author} {\bibfnamefont {A.}~\bibnamefont {Karch}},
  \bibinfo {author} {\bibfnamefont {C.}~\bibnamefont {Perez-Pardavila}},
  \bibinfo {author} {\bibfnamefont {S.}~\bibnamefont {Raju}}, \bibinfo {author}
  {\bibfnamefont {L.}~\bibnamefont {Randall}}, \bibinfo {author} {\bibfnamefont
  {M.}~\bibnamefont {Riojas}}, \ and\ \bibinfo {author} {\bibfnamefont
  {S.}~\bibnamefont {Shashi}},\ }\href@noop {} {\  (\bibinfo {year} {2020})},\
  \Eprint {http://arxiv.org/abs/2012.04671} {arXiv:2012.04671 [hep-th]}
  \BibitemShut {NoStop}%
\bibitem [{\citenamefont {Geng}\ \emph
  {et~al.}(2021{\natexlab{a}})\citenamefont {Geng}, \citenamefont {Karch},
  \citenamefont {Perez-Pardavila}, \citenamefont {Raju}, \citenamefont
  {Randall}, \citenamefont {Riojas},\ and\ \citenamefont
  {Shashi}}]{Geng:2021hlu}%
  \BibitemOpen
  \bibfield  {author} {\bibinfo {author} {\bibfnamefont {H.}~\bibnamefont
  {Geng}}, \bibinfo {author} {\bibfnamefont {A.}~\bibnamefont {Karch}},
  \bibinfo {author} {\bibfnamefont {C.}~\bibnamefont {Perez-Pardavila}},
  \bibinfo {author} {\bibfnamefont {S.}~\bibnamefont {Raju}}, \bibinfo {author}
  {\bibfnamefont {L.}~\bibnamefont {Randall}}, \bibinfo {author} {\bibfnamefont
  {M.}~\bibnamefont {Riojas}}, \ and\ \bibinfo {author} {\bibfnamefont
  {S.}~\bibnamefont {Shashi}},\ }\href@noop {} {\  (\bibinfo {year}
  {2021}{\natexlab{a}})},\ \Eprint {http://arxiv.org/abs/2107.03390}
  {arXiv:2107.03390 [hep-th]} \BibitemShut {NoStop}%
\bibitem [{\citenamefont {Geng}\ \emph
  {et~al.}(2021{\natexlab{b}})\citenamefont {Geng}, \citenamefont {Karch},
  \citenamefont {Perez-Pardavila}, \citenamefont {Raju}, \citenamefont
  {Randall}, \citenamefont {Riojas},\ and\ \citenamefont
  {Shashi}}]{Geng:2021mic}%
  \BibitemOpen
  \bibfield  {author} {\bibinfo {author} {\bibfnamefont {H.}~\bibnamefont
  {Geng}}, \bibinfo {author} {\bibfnamefont {A.}~\bibnamefont {Karch}},
  \bibinfo {author} {\bibfnamefont {C.}~\bibnamefont {Perez-Pardavila}},
  \bibinfo {author} {\bibfnamefont {S.}~\bibnamefont {Raju}}, \bibinfo {author}
  {\bibfnamefont {L.}~\bibnamefont {Randall}}, \bibinfo {author} {\bibfnamefont
  {M.}~\bibnamefont {Riojas}}, \ and\ \bibinfo {author} {\bibfnamefont
  {S.}~\bibnamefont {Shashi}},\ }\href@noop {} {\  (\bibinfo {year}
  {2021}{\natexlab{b}})},\ \Eprint {http://arxiv.org/abs/2112.09132}
  {arXiv:2112.09132 [hep-th]} \BibitemShut {NoStop}%
\bibitem [{\citenamefont {Geng}\ \emph {et~al.}()\citenamefont {Geng},
  \citenamefont {Randall},\ and\ \citenamefont {Swanson}}]{Geng:2022yyy}%
  \BibitemOpen
  \bibfield  {author} {\bibinfo {author} {\bibfnamefont {H.}~\bibnamefont
  {Geng}}, \bibinfo {author} {\bibfnamefont {L.}~\bibnamefont {Randall}}, \
  and\ \bibinfo {author} {\bibfnamefont {E.}~\bibnamefont {Swanson}},\
  }\href@noop {} {\bibinfo  {journal} {To appear}\ }\BibitemShut {NoStop}%
\bibitem [{\citenamefont {Akal}\ \emph {et~al.}(2020)\citenamefont {Akal},
  \citenamefont {Kusuki}, \citenamefont {Takayanagi},\ and\ \citenamefont
  {Wei}}]{Akal:2020wfl}%
  \BibitemOpen
\bibfield  {journal} {  }\bibfield  {author} {\bibinfo {author} {\bibfnamefont
  {I.}~\bibnamefont {Akal}}, \bibinfo {author} {\bibfnamefont {Y.}~\bibnamefont
  {Kusuki}}, \bibinfo {author} {\bibfnamefont {T.}~\bibnamefont {Takayanagi}},
  \ and\ \bibinfo {author} {\bibfnamefont {Z.}~\bibnamefont {Wei}},\
  }\href@noop {} {\  (\bibinfo {year} {2020})},\ \Eprint
  {http://arxiv.org/abs/2007.06800} {arXiv:2007.06800 [hep-th]} \BibitemShut
  {NoStop}%
\bibitem [{\citenamefont {Miao}(2020)}]{Miao:2020oey}%
  \BibitemOpen
  \bibfield  {author} {\bibinfo {author} {\bibfnamefont {R.-X.}\ \bibnamefont
  {Miao}},\ }\href@noop {} {\  (\bibinfo {year} {2020})},\ \Eprint
  {http://arxiv.org/abs/2009.06263} {arXiv:2009.06263 [hep-th]} \BibitemShut
  {NoStop}%
\bibitem [{\citenamefont {Maldacena}(1999)}]{Maldacena:1997re}%
  \BibitemOpen
  \bibfield  {author} {\bibinfo {author} {\bibfnamefont {J.~M.}\ \bibnamefont
  {Maldacena}},\ }\href {\doibase 10.1023/A:1026654312961,
  10.4310/ATMP.1998.v2.n2.a1} {\bibfield  {journal} {\bibinfo  {journal} {Int.
  J. Theor. Phys.}\ }\textbf {\bibinfo {volume} {38}},\ \bibinfo {pages} {1113}
  (\bibinfo {year} {1999})},\ \bibinfo {note} {[Adv. Theor. Math.
  Phys.2,231(1998)]},\ \Eprint {http://arxiv.org/abs/hep-th/9711200}
  {arXiv:hep-th/9711200 [hep-th]} \BibitemShut {NoStop}%
\bibitem [{\citenamefont {Witten}(1998)}]{Witten:1998qj}%
  \BibitemOpen
  \bibfield  {author} {\bibinfo {author} {\bibfnamefont {E.}~\bibnamefont
  {Witten}},\ }\href {\doibase 10.4310/ATMP.1998.v2.n2.a2} {\bibfield
  {journal} {\bibinfo  {journal} {Adv. Theor. Math. Phys.}\ }\textbf {\bibinfo
  {volume} {2}},\ \bibinfo {pages} {253} (\bibinfo {year} {1998})},\ \Eprint
  {http://arxiv.org/abs/hep-th/9802150} {arXiv:hep-th/9802150 [hep-th]}
  \BibitemShut {NoStop}%
\bibitem [{\citenamefont {Gubser}\ \emph {et~al.}(1998)\citenamefont {Gubser},
  \citenamefont {Klebanov},\ and\ \citenamefont {Polyakov}}]{Gubser:1998bc}%
  \BibitemOpen
  \bibfield  {author} {\bibinfo {author} {\bibfnamefont {S.~S.}\ \bibnamefont
  {Gubser}}, \bibinfo {author} {\bibfnamefont {I.~R.}\ \bibnamefont
  {Klebanov}}, \ and\ \bibinfo {author} {\bibfnamefont {A.~M.}\ \bibnamefont
  {Polyakov}},\ }\href {\doibase 10.1016/S0370-2693(98)00377-3} {\bibfield
  {journal} {\bibinfo  {journal} {Phys. Lett.}\ }\textbf {\bibinfo {volume}
  {B428}},\ \bibinfo {pages} {105} (\bibinfo {year} {1998})},\ \Eprint
  {http://arxiv.org/abs/hep-th/9802109} {arXiv:hep-th/9802109 [hep-th]}
  \BibitemShut {NoStop}%
\bibitem [{\citenamefont {Randall}\ and\ \citenamefont
  {Sundrum}(1999{\natexlab{a}})}]{Randall:1999ee}%
  \BibitemOpen
  \bibfield  {author} {\bibinfo {author} {\bibfnamefont {L.}~\bibnamefont
  {Randall}}\ and\ \bibinfo {author} {\bibfnamefont {R.}~\bibnamefont
  {Sundrum}},\ }\href {\doibase 10.1103/PhysRevLett.83.3370} {\bibfield
  {journal} {\bibinfo  {journal} {Phys. Rev. Lett.}\ }\textbf {\bibinfo
  {volume} {83}},\ \bibinfo {pages} {3370} (\bibinfo {year}
  {1999}{\natexlab{a}})},\ \Eprint {http://arxiv.org/abs/hep-ph/9905221}
  {arXiv:hep-ph/9905221} \BibitemShut {NoStop}%
\bibitem [{\citenamefont {Jackiw}(1985)}]{Jackiw:1984je}%
  \BibitemOpen
  \bibfield  {author} {\bibinfo {author} {\bibfnamefont {R.}~\bibnamefont
  {Jackiw}},\ }\href {\doibase 10.1016/0550-3213(85)90448-1} {\bibfield
  {journal} {\bibinfo  {journal} {Nucl. Phys. B}\ }\textbf {\bibinfo {volume}
  {252}},\ \bibinfo {pages} {343} (\bibinfo {year} {1985})}\BibitemShut
  {NoStop}%
\bibitem [{\citenamefont {Teitelboim}(1983)}]{Teitelboim:1983ux}%
  \BibitemOpen
  \bibfield  {author} {\bibinfo {author} {\bibfnamefont {C.}~\bibnamefont
  {Teitelboim}},\ }\href {\doibase 10.1016/0370-2693(83)90012-6} {\bibfield
  {journal} {\bibinfo  {journal} {Phys. Lett. B}\ }\textbf {\bibinfo {volume}
  {126}},\ \bibinfo {pages} {41} (\bibinfo {year} {1983})}\BibitemShut
  {NoStop}%
\bibitem [{\citenamefont {Dong}\ \emph {et~al.}(2016)\citenamefont {Dong},
  \citenamefont {Harlow},\ and\ \citenamefont {Wall}}]{Dong:2016eik}%
  \BibitemOpen
  \bibfield  {author} {\bibinfo {author} {\bibfnamefont {X.}~\bibnamefont
  {Dong}}, \bibinfo {author} {\bibfnamefont {D.}~\bibnamefont {Harlow}}, \ and\
  \bibinfo {author} {\bibfnamefont {A.~C.}\ \bibnamefont {Wall}},\ }\href
  {\doibase 10.1103/PhysRevLett.117.021601} {\bibfield  {journal} {\bibinfo
  {journal} {Phys. Rev. Lett.}\ }\textbf {\bibinfo {volume} {117}},\ \bibinfo
  {pages} {021601} (\bibinfo {year} {2016})},\ \Eprint
  {http://arxiv.org/abs/1601.05416} {arXiv:1601.05416 [hep-th]} \BibitemShut
  {NoStop}%
\bibitem [{\citenamefont {Almheiri}\ and\ \citenamefont
  {Polchinski}(2015)}]{Almheiri:2014cka}%
  \BibitemOpen
  \bibfield  {author} {\bibinfo {author} {\bibfnamefont {A.}~\bibnamefont
  {Almheiri}}\ and\ \bibinfo {author} {\bibfnamefont {J.}~\bibnamefont
  {Polchinski}},\ }\href {\doibase 10.1007/JHEP11(2015)014} {\bibfield
  {journal} {\bibinfo  {journal} {JHEP}\ }\textbf {\bibinfo {volume} {11}},\
  \bibinfo {pages} {014} (\bibinfo {year} {2015})},\ \Eprint
  {http://arxiv.org/abs/1402.6334} {arXiv:1402.6334 [hep-th]} \BibitemShut
  {NoStop}%
\bibitem [{\citenamefont {Jensen}(2016)}]{Jensen:2016pah}%
  \BibitemOpen
  \bibfield  {author} {\bibinfo {author} {\bibfnamefont {K.}~\bibnamefont
  {Jensen}},\ }\href {\doibase 10.1103/PhysRevLett.117.111601} {\bibfield
  {journal} {\bibinfo  {journal} {Phys. Rev. Lett.}\ }\textbf {\bibinfo
  {volume} {117}},\ \bibinfo {pages} {111601} (\bibinfo {year} {2016})},\
  \Eprint {http://arxiv.org/abs/1605.06098} {arXiv:1605.06098 [hep-th]}
  \BibitemShut {NoStop}%
\bibitem [{\citenamefont {Maldacena}\ \emph {et~al.}(2016)\citenamefont
  {Maldacena}, \citenamefont {Stanford},\ and\ \citenamefont
  {Yang}}]{Maldacena:2016upp}%
  \BibitemOpen
  \bibfield  {author} {\bibinfo {author} {\bibfnamefont {J.}~\bibnamefont
  {Maldacena}}, \bibinfo {author} {\bibfnamefont {D.}~\bibnamefont {Stanford}},
  \ and\ \bibinfo {author} {\bibfnamefont {Z.}~\bibnamefont {Yang}},\ }\href
  {\doibase 10.1093/ptep/ptw124} {\bibfield  {journal} {\bibinfo  {journal}
  {PTEP}\ }\textbf {\bibinfo {volume} {2016}},\ \bibinfo {pages} {12C104}
  (\bibinfo {year} {2016})},\ \Eprint {http://arxiv.org/abs/1606.01857}
  {arXiv:1606.01857 [hep-th]} \BibitemShut {NoStop}%
\bibitem [{\citenamefont {Engels\"oy}\ \emph {et~al.}(2016)\citenamefont
  {Engels\"oy}, \citenamefont {Mertens},\ and\ \citenamefont
  {Verlinde}}]{Engelsoy:2016xyb}%
  \BibitemOpen
  \bibfield  {author} {\bibinfo {author} {\bibfnamefont {J.}~\bibnamefont
  {Engels\"oy}}, \bibinfo {author} {\bibfnamefont {T.~G.}\ \bibnamefont
  {Mertens}}, \ and\ \bibinfo {author} {\bibfnamefont {H.}~\bibnamefont
  {Verlinde}},\ }\href {\doibase 10.1007/JHEP07(2016)139} {\bibfield  {journal}
  {\bibinfo  {journal} {JHEP}\ }\textbf {\bibinfo {volume} {07}},\ \bibinfo
  {pages} {139} (\bibinfo {year} {2016})},\ \Eprint
  {http://arxiv.org/abs/1606.03438} {arXiv:1606.03438 [hep-th]} \BibitemShut
  {NoStop}%
\bibitem [{\citenamefont {Saad}\ \emph {et~al.}(2019)\citenamefont {Saad},
  \citenamefont {Shenker},\ and\ \citenamefont {Stanford}}]{Saad:2019lba}%
  \BibitemOpen
  \bibfield  {author} {\bibinfo {author} {\bibfnamefont {P.}~\bibnamefont
  {Saad}}, \bibinfo {author} {\bibfnamefont {S.~H.}\ \bibnamefont {Shenker}}, \
  and\ \bibinfo {author} {\bibfnamefont {D.}~\bibnamefont {Stanford}},\
  }\href@noop {} {\  (\bibinfo {year} {2019})},\ \Eprint
  {http://arxiv.org/abs/1903.11115} {arXiv:1903.11115 [hep-th]} \BibitemShut
  {NoStop}%
\bibitem [{\citenamefont {Geng}()}]{Geng:2022xxx}%
  \BibitemOpen
  \bibfield  {author} {\bibinfo {author} {\bibfnamefont {H.}~\bibnamefont
  {Geng}},\ }\href@noop {} {\bibinfo  {journal} {To appear}\ }\BibitemShut
  {NoStop}%
\bibitem [{\citenamefont {Randall}\ and\ \citenamefont
  {Sundrum}(1999{\natexlab{b}})}]{Randall:1999vf}%
  \BibitemOpen
\bibfield  {journal} {  }\bibfield  {author} {\bibinfo {author} {\bibfnamefont
  {L.}~\bibnamefont {Randall}}\ and\ \bibinfo {author} {\bibfnamefont
  {R.}~\bibnamefont {Sundrum}},\ }\href {\doibase 10.1103/PhysRevLett.83.4690}
  {\bibfield  {journal} {\bibinfo  {journal} {Phys. Rev. Lett.}\ }\textbf
  {\bibinfo {volume} {83}},\ \bibinfo {pages} {4690} (\bibinfo {year}
  {1999}{\natexlab{b}})},\ \Eprint {http://arxiv.org/abs/hep-th/9906064}
  {arXiv:hep-th/9906064} \BibitemShut {NoStop}%
\bibitem [{\citenamefont {Fujita}\ \emph {et~al.}(2011)\citenamefont {Fujita},
  \citenamefont {Takayanagi},\ and\ \citenamefont {Tonni}}]{Fujita:2011fp}%
  \BibitemOpen
  \bibfield  {author} {\bibinfo {author} {\bibfnamefont {M.}~\bibnamefont
  {Fujita}}, \bibinfo {author} {\bibfnamefont {T.}~\bibnamefont {Takayanagi}},
  \ and\ \bibinfo {author} {\bibfnamefont {E.}~\bibnamefont {Tonni}},\ }\href
  {\doibase 10.1007/JHEP11(2011)043} {\bibfield  {journal} {\bibinfo  {journal}
  {JHEP}\ }\textbf {\bibinfo {volume} {11}},\ \bibinfo {pages} {043} (\bibinfo
  {year} {2011})},\ \Eprint {http://arxiv.org/abs/1108.5152} {arXiv:1108.5152
  [hep-th]} \BibitemShut {NoStop}%
\bibitem [{\citenamefont {Karch}\ and\ \citenamefont
  {Randall}(2020)}]{Karch:2020iit}%
  \BibitemOpen
  \bibfield  {author} {\bibinfo {author} {\bibfnamefont {A.}~\bibnamefont
  {Karch}}\ and\ \bibinfo {author} {\bibfnamefont {L.}~\bibnamefont
  {Randall}},\ }\href {\doibase 10.1007/JHEP09(2020)166} {\bibfield  {journal}
  {\bibinfo  {journal} {JHEP}\ }\textbf {\bibinfo {volume} {09}},\ \bibinfo
  {pages} {166} (\bibinfo {year} {2020})},\ \Eprint
  {http://arxiv.org/abs/2006.10061} {arXiv:2006.10061 [hep-th]} \BibitemShut
  {NoStop}%
\bibitem [{\citenamefont {Polchinski}(2007)}]{Polchinski:1998rq}%
  \BibitemOpen
  \bibfield  {author} {\bibinfo {author} {\bibfnamefont {J.}~\bibnamefont
  {Polchinski}},\ }\href {\doibase 10.1017/CBO9780511816079} {\emph {\bibinfo
  {title} {{String theory. Vol. 1}}}},\ Cambridge Monographs on Mathematical
  Physics\ (\bibinfo  {publisher} {Cambridge University Press},\ \bibinfo
  {year} {2007})\BibitemShut {NoStop}%
\bibitem [{\citenamefont {Johnson}(2005)}]{Johnson:2003gi}%
  \BibitemOpen
  \bibfield  {author} {\bibinfo {author} {\bibfnamefont {C.~V.}\ \bibnamefont
  {Johnson}},\ }\href {\doibase 10.1017/CBO9780511606540} {\emph {\bibinfo
  {title} {{D-branes}}}},\ Cambridge Monographs on Mathematical Physics\
  (\bibinfo  {publisher} {Cambridge University Press},\ \bibinfo {year}
  {2005})\BibitemShut {NoStop}%
\bibitem [{\citenamefont {Arkani-Hamed}\ \emph {et~al.}(2001)\citenamefont
  {Arkani-Hamed}, \citenamefont {Porrati},\ and\ \citenamefont
  {Randall}}]{ArkaniHamed:2000ds}%
  \BibitemOpen
  \bibfield  {author} {\bibinfo {author} {\bibfnamefont {N.}~\bibnamefont
  {Arkani-Hamed}}, \bibinfo {author} {\bibfnamefont {M.}~\bibnamefont
  {Porrati}}, \ and\ \bibinfo {author} {\bibfnamefont {L.}~\bibnamefont
  {Randall}},\ }\href {\doibase 10.1088/1126-6708/2001/08/017} {\bibfield
  {journal} {\bibinfo  {journal} {JHEP}\ }\textbf {\bibinfo {volume} {08}},\
  \bibinfo {pages} {017} (\bibinfo {year} {2001})},\ \Eprint
  {http://arxiv.org/abs/hep-th/0012148} {arXiv:hep-th/0012148} \BibitemShut
  {NoStop}%
\bibitem [{\citenamefont {Maldacena}\ \emph {et~al.}(1999)\citenamefont
  {Maldacena}, \citenamefont {Michelson},\ and\ \citenamefont
  {Strominger}}]{Maldacena:1998uz}%
  \BibitemOpen
  \bibfield  {author} {\bibinfo {author} {\bibfnamefont {J.~M.}\ \bibnamefont
  {Maldacena}}, \bibinfo {author} {\bibfnamefont {J.}~\bibnamefont
  {Michelson}}, \ and\ \bibinfo {author} {\bibfnamefont {A.}~\bibnamefont
  {Strominger}},\ }\href {\doibase 10.1088/1126-6708/1999/02/011} {\bibfield
  {journal} {\bibinfo  {journal} {JHEP}\ }\textbf {\bibinfo {volume} {02}},\
  \bibinfo {pages} {011} (\bibinfo {year} {1999})},\ \Eprint
  {http://arxiv.org/abs/hep-th/9812073} {arXiv:hep-th/9812073} \BibitemShut
  {NoStop}%
\bibitem [{\citenamefont {Harlow}\ and\ \citenamefont
  {Jafferis}(2020)}]{Harlow:2018tqv}%
  \BibitemOpen
  \bibfield  {author} {\bibinfo {author} {\bibfnamefont {D.}~\bibnamefont
  {Harlow}}\ and\ \bibinfo {author} {\bibfnamefont {D.}~\bibnamefont
  {Jafferis}},\ }\href {\doibase 10.1007/JHEP02(2020)177} {\bibfield  {journal}
  {\bibinfo  {journal} {JHEP}\ }\textbf {\bibinfo {volume} {02}},\ \bibinfo
  {pages} {177} (\bibinfo {year} {2020})},\ \Eprint
  {http://arxiv.org/abs/1804.01081} {arXiv:1804.01081 [hep-th]} \BibitemShut
  {NoStop}%
\bibitem [{\citenamefont {Nayak}\ \emph {et~al.}(2018)\citenamefont {Nayak},
  \citenamefont {Shukla}, \citenamefont {Soni}, \citenamefont {Trivedi},\ and\
  \citenamefont {Vishal}}]{Nayak:2018qej}%
  \BibitemOpen
  \bibfield  {author} {\bibinfo {author} {\bibfnamefont {P.}~\bibnamefont
  {Nayak}}, \bibinfo {author} {\bibfnamefont {A.}~\bibnamefont {Shukla}},
  \bibinfo {author} {\bibfnamefont {R.~M.}\ \bibnamefont {Soni}}, \bibinfo
  {author} {\bibfnamefont {S.~P.}\ \bibnamefont {Trivedi}}, \ and\ \bibinfo
  {author} {\bibfnamefont {V.}~\bibnamefont {Vishal}},\ }\href {\doibase
  10.1007/JHEP09(2018)048} {\bibfield  {journal} {\bibinfo  {journal} {JHEP}\
  }\textbf {\bibinfo {volume} {09}},\ \bibinfo {pages} {048} (\bibinfo {year}
  {2018})},\ \Eprint {http://arxiv.org/abs/1802.09547} {arXiv:1802.09547
  [hep-th]} \BibitemShut {NoStop}%
\bibitem [{\citenamefont {Yang}(2019)}]{Yang:2018gdb}%
  \BibitemOpen
  \bibfield  {author} {\bibinfo {author} {\bibfnamefont {Z.}~\bibnamefont
  {Yang}},\ }\href {\doibase 10.1007/JHEP05(2019)205} {\bibfield  {journal}
  {\bibinfo  {journal} {JHEP}\ }\textbf {\bibinfo {volume} {05}},\ \bibinfo
  {pages} {205} (\bibinfo {year} {2019})},\ \Eprint
  {http://arxiv.org/abs/1809.08647} {arXiv:1809.08647 [hep-th]} \BibitemShut
  {NoStop}%
\bibitem [{\citenamefont {Ryu}\ and\ \citenamefont
  {Takayanagi}(2006{\natexlab{a}})}]{Ryu:2006bv}%
  \BibitemOpen
  \bibfield  {author} {\bibinfo {author} {\bibfnamefont {S.}~\bibnamefont
  {Ryu}}\ and\ \bibinfo {author} {\bibfnamefont {T.}~\bibnamefont
  {Takayanagi}},\ }\href {\doibase 10.1103/PhysRevLett.96.181602} {\bibfield
  {journal} {\bibinfo  {journal} {Phys. Rev. Lett.}\ }\textbf {\bibinfo
  {volume} {96}},\ \bibinfo {pages} {181602} (\bibinfo {year}
  {2006}{\natexlab{a}})},\ \Eprint {http://arxiv.org/abs/hep-th/0603001}
  {arXiv:hep-th/0603001 [hep-th]} \BibitemShut {NoStop}%
\bibitem [{\citenamefont {Ryu}\ and\ \citenamefont
  {Takayanagi}(2006{\natexlab{b}})}]{Ryu:2006ef}%
  \BibitemOpen
  \bibfield  {author} {\bibinfo {author} {\bibfnamefont {S.}~\bibnamefont
  {Ryu}}\ and\ \bibinfo {author} {\bibfnamefont {T.}~\bibnamefont
  {Takayanagi}},\ }\href {\doibase 10.1088/1126-6708/2006/08/045} {\bibfield
  {journal} {\bibinfo  {journal} {JHEP}\ }\textbf {\bibinfo {volume} {08}},\
  \bibinfo {pages} {045} (\bibinfo {year} {2006}{\natexlab{b}})},\ \Eprint
  {http://arxiv.org/abs/hep-th/0605073} {arXiv:hep-th/0605073 [hep-th]}
  \BibitemShut {NoStop}%
\bibitem [{\citenamefont {Jafferis}\ \emph {et~al.}(2016)\citenamefont
  {Jafferis}, \citenamefont {Lewkowycz}, \citenamefont {Maldacena},\ and\
  \citenamefont {Suh}}]{Jafferis:2015del}%
  \BibitemOpen
  \bibfield  {author} {\bibinfo {author} {\bibfnamefont {D.~L.}\ \bibnamefont
  {Jafferis}}, \bibinfo {author} {\bibfnamefont {A.}~\bibnamefont {Lewkowycz}},
  \bibinfo {author} {\bibfnamefont {J.}~\bibnamefont {Maldacena}}, \ and\
  \bibinfo {author} {\bibfnamefont {S.~J.}\ \bibnamefont {Suh}},\ }\href
  {\doibase 10.1007/JHEP06(2016)004} {\bibfield  {journal} {\bibinfo  {journal}
  {JHEP}\ }\textbf {\bibinfo {volume} {06}},\ \bibinfo {pages} {004} (\bibinfo
  {year} {2016})},\ \Eprint {http://arxiv.org/abs/1512.06431} {arXiv:1512.06431
  [hep-th]} \BibitemShut {NoStop}%
\bibitem [{\citenamefont {Hubeny}\ \emph {et~al.}(2007)\citenamefont {Hubeny},
  \citenamefont {Rangamani},\ and\ \citenamefont {Takayanagi}}]{Hubeny:2007xt}%
  \BibitemOpen
  \bibfield  {author} {\bibinfo {author} {\bibfnamefont {V.~E.}\ \bibnamefont
  {Hubeny}}, \bibinfo {author} {\bibfnamefont {M.}~\bibnamefont {Rangamani}}, \
  and\ \bibinfo {author} {\bibfnamefont {T.}~\bibnamefont {Takayanagi}},\
  }\href {\doibase 10.1088/1126-6708/2007/07/062} {\bibfield  {journal}
  {\bibinfo  {journal} {JHEP}\ }\textbf {\bibinfo {volume} {07}},\ \bibinfo
  {pages} {062} (\bibinfo {year} {2007})},\ \Eprint
  {http://arxiv.org/abs/0705.0016} {arXiv:0705.0016 [hep-th]} \BibitemShut
  {NoStop}%
\bibitem [{\citenamefont {Wall}(2014)}]{Wall:2012uf}%
  \BibitemOpen
  \bibfield  {author} {\bibinfo {author} {\bibfnamefont {A.~C.}\ \bibnamefont
  {Wall}},\ }\href {\doibase 10.1088/0264-9381/31/22/225007} {\bibfield
  {journal} {\bibinfo  {journal} {Class. Quant. Grav.}\ }\textbf {\bibinfo
  {volume} {31}},\ \bibinfo {pages} {225007} (\bibinfo {year} {2014})},\
  \Eprint {http://arxiv.org/abs/1211.3494} {arXiv:1211.3494 [hep-th]}
  \BibitemShut {NoStop}%
\end{thebibliography}%
\end{document}